%% file: klcf.tex
\documentclass{llncs}

\usepackage{xspace,amsmath,url}
\usepackage{color}
\usepackage{booktabs}

\newcommand{\exclude}[1]{}

\newcounter{lnoc}

\newcommand{\lno}[1][0]{{\footnotesize\sffamily 
\ifnum#1=0
\stepcounter{lnoc} 
\ifnum\thelnoc<10
\phantom0%
\fi
\thelnoc
\else
\thelnoc.#1
\fi
}\>}

\begin{document}

\title{A note on the longest common substring with $k$-mismatches problem}

\author{Szymon Grabowski}
\institute{Lodz University of Technology, Institute of Applied Computer Science,\\
  Al.\ Politechniki 11, 90--924 {\L}\'od\'z, Poland, 
  \email{sgrabow@kis.p.lodz.pl}
}

\maketitle

\begin{abstract}
The recently introduced longest common substring 
with $k$-mismatches ($k$-LCF) problem
is to find, given two sequences $S_1$ and $S_2$ of length $n$ each, 
a longest substring $A_1$ of $S_1$ and $A_2$ of $S_2$ such that the 
Hamming distance between $A_1$ and $A_2$ is at most $k$.
So far, the only subquadratic time result for this problem was known 
for $k = 1$~\cite{FGKU2014}.
We first present two output-dependent algorithms solving the $k$-LCF problem 
and show that for 
$k = O(\log^{1-\varepsilon} n)$, where $\varepsilon > 0$,
at least one of them 
works in subquadratic time, using $O(n)$ words of space.
The choice of one of these two algorithms to be applied for a given input 
can be done after linear time and space preprocessing.
Finally we present a tabulation-based algorithm working, 
in its range of applicability, in 
$O(n^2\log\min(k+\ell_0, \sigma)/\log n)$ time, where 
$\ell_0$ is the length of the standard longest common substring.
\end{abstract}

\input{intro.tex}
\input{alg.tex}
\input{concs.tex}

\bibliographystyle{abbrv}
\bibliography{klcf}

\end{document}

%% file: intro.tex
\section{Introduction}
The longest common substring (or factor) problem (LCF) is 
to find the longest contiguous string shared by two strings $S_1$ and $S_2$, 
of length $n$ and $m$, $m \leq n$, respectively.
W.l.o.g.
%%\footnote{Might be not true for very small $m$...} 
(and to simplify notation) we assume $n = m$.
A generalization of this problem allows for approximate matches, namely 
in the Hamming distance sense.

Formally, we define the {\em longest common substring with k mismatches} ($k$-LCF) 
as follows.
Given two strings, $S_1[1 \ldots n]$ and $S_2[1 \ldots n]$, 
over an integer alphabet $\Sigma$ of size $\sigma$,
and integer $k$, 
find a pair of strings $S_1[i_1 \ldots i_1+\ell-1]$ and $S_2[i_2 \ldots i_2+\ell-1]$ 
such that $S_1[i_1 \ldots i_1+\ell-1]$ and $S_2[i_2 \ldots i_2+\ell-1]$ differ
in at most $k$ positions 
(i.e., $|\{i_1 \leq pos \leq i_1+\ell-1: S_1[pos] \neq S_2[pos + (i_2 - i_1)] \}| \leq k$)
and the string length $\ell$ is maximized.
For simplicity, let us further assume that the two considered substrings differ 
in exactly $k$ positions.

Following~\cite{FGKU2014}, let $\phi(i, j)$ be the length of the longest suffix 
of $S_1[1 \ldots i]$ and $S_2[1 \ldots j]$ matching with up to $k$ mismatches.
A match with up to $k$ mismatches will sometimes be called a $k$-approximate match.
To simplify notation, we will 
use the symbols $\ell_k = |\operatorname{{\it k}-LCF}(S_1, S_2)|$ and 
$\ell_0 = |\operatorname{LCF}(S_1, S_2)|$ throughout the paper.
Whenever 
clear 
from the context, we will talk about simply 
a common substring (or a longest common substring) in the 
$k$-LCF sense.
The popcount function $popc(B)$ returns the number of 1s in bit-vector $B$.
We assume a machine word of length $w = \Theta(\log n)$ bits.
All logarithms are in base 2.

The problem history is short.
Babenko and Starikovskaya~\cite{BS2011} gave an $O(n^2)$-time, $O(n)$-space 
algorithm for 1-LCF.
Flouri et al.~\cite{FGKU2014} recently presented two algorithms: 
an $O(n\log n)$-time, $O(n)$-space algorithm for 1-LCF 
and an $O(n^2)$-time, $O(1)$-space one for the general $k$-LCF.
They also gave a variant of the second algorithm, involving longest 
common extension queries (LCE), with $O(n + |\mathcal{K}|)$ time and $O(n)$ space, 
where $\mathcal{K}$ is the set of all mismatching pairs of symbols from $S_1$ and $S_2$, 
i.e., $\mathcal{K} = \{(i, j): S_1[i] \neq S_2[j]\}$ (the $\mathcal{K}$ definition 
in the cited paper is slightly different, yet it does not matter for the presented 
complexity).

%% file: alg.tex
\section{Our algorithms}\label{sec:our_alg}
%%%%%%%%%%%%%%%%%%%%%%%%%%%%%%%%%%%%%%%%%%%%%%%%%%%%
In the subsections to follow we are going to present three algorithms.
They make use of the length of the (standard) longest common substring 
of $S_1$ and $S_2$.
In a preliminary step thus we compute $\ell_0$ in linear time and space, 
using the classical method~\cite{Gus1997}.
Note that $\ell_0 + k \leq \ell_k \leq (k+1) \ell_0 + k$.
The found value of $\ell_0$ gives thus some bounds on the unknown value of $\ell_k$.
Further on we assume that $\ell_0 > 0$, otherwise we trivially obtain $\ell_k = k$ 
(in this extreme case, any substring of $S_1$ of length $k$ is an $k$-LCF of $S_1$ and $S_2$).
%% The third algorithm is based on tabulation.
In the exposition, we focus on finding the $k$-LCF length, but it is obvious from 
the corresponding descriptions that the desired substring and its location 
is $S_1$ and $S_2$ are found too in the same time complexity.

\subsection{Neighborhood generation based algorithm} \label{sec:ng}
%%%%%%%%%%%%%%%%%%%%%%%%%%%%%%%%%%%%%%%%%%%%%%%%%%%%
Assume that $(k+1) \ell_0 + k$ is small enough.
We are going to find the smallest $j \in \mathcal{I}$, 
where $\mathcal{I} = \{\ell_0 + k + 1, \ell_0 + k + 2, \ldots, (k+1) \ell_0 + k + 1\}$, 
such that there is no substring from $S_1$ of length $j$ that occurs
with at most $k$ mismatches in $S_2$. 
This will mean that $\ell_k = j - 1$.

To check if $S_1$ and $S_2$ have a $k$-approximate match of length $j$ 
we generate the explicit neighborhood of each substring of length $j$ 
from $S_1$ and $S_2$, deleting $k$ symbols at all possible subsets of $k$ positions.
If two strings, one from $S_1$ and the other from $S_2$, are equal after such a deletion 
and the deleted symbols' position subsets are also equal, we have a 
$k$-approximate match.
Let us give an example: if the string is \texttt{abbac} and $k=2$, 
then the neighborhood is: 
\texttt{abb} (4, 5),
\texttt{aba} (3, 5),
\texttt{abc} (3, 4),
\texttt{aba} (2, 5),
\texttt{abc} (2, 4),
\texttt{aac} (2, 3),
\texttt{bba} (1, 5),
\texttt{bbc} (1, 4),
\texttt{bac} (1, 3),
\texttt{bac} (1, 2). 
This technique was invented by Mor and Frankel~\cite{MF1982} for dictionary matching with 
one error and then generalized to $k$ errors by Bocek et al.~\cite{BHSH2007}.
Let us call the original string (\texttt{abbac} in the example) a source 
for the generated strings (keywords) from the neighborhood.
We will store each keyword using $O(k)$ space: 
the position of its source in $S_1\#S_2$ and the $k$ delete positions.
All comparisons between such strings, whether to check for a match or to settle
their lexicographical order in a sorting phase, will take $O(k)$ time, 
thanks to using LCE queries.

The neighborhood of a string of length $j$ is of size $O(j^k)$ (keywords) 
and is generated using $O(k j^k)$ time and space. 
For the whole sequences, having $n-j+1$ source strings each, 
the total neighborhood generation time and space is thus $O(n k j^k)$.
We sort the resulting collection of $N = O(n j^k)$ keywords 
from $S_1$ in $O(N k\log N)$ time, obtaining a keyword index,  
and binary-search for each of the resulting $N$ keywords from $S_2$ in the built index.
The search phase takes $O(N k\log N)$ time as well.
This means that testing for an existence of a $k$-approximate match of 
specified length $j$ between $S_1$ and $S_2$ 
takes $O(N k\log N) = O(n k j^k (\log n + k\log j))$ time 
and needs $O(n k j^k)$ space.
As we need to examine $O(\log |\mathcal{I}|)$ values of $j$ from $\mathcal{I}$, 
in a binary search manner, 
the total time complexity becomes 
\begin{equation} \label{eq:first}
O(nk((k+1)(\ell_0 + 1))^k (\log n + k\log \ell_0 + k\log k)(\log k + \log\ell_0)).
\end{equation}
As each value of $j$ is processed separately, the peak space use 
corresponds to the largest inspected $j$ and the space complexity becomes 
$O(nk((k+1)(\ell_0 + 1))^k)$.

Note that even for constant values of $k$ and $\ell_0$ the space use may 
be prohibitive.
To reduce the space use, we partition $S_1$ $h$ into equal-length pieces 
and for each of them in turn build a sorted array of all generated keywords, 
and query all keywords generated from $S_2$ with this index.
This seemingly makes the time grow by factor $O(h)$ (as each substring constructed 
from $S_2$ is queried $h$ times in total) and the space reduce by factor $O(h)$.
A more careful look however reveals that 
successive pieces of $S_1$ must have an overlap 
of size $j$, not to miss any match.
Another (minor) change is that the binary searches are performed over a collection 
(close to) $h$ times smaller, which reduces the corresponding log-factor.

Let us consider the space complexity. 
Instead of $O(nk((k+1)(\ell_0 + 1))^k)$ we now have 
$O( nk((k+1)(\ell_0 + 1))^k / h + hk((k+1)(\ell_0 + 1))^{k+1} )$ space, 
where the second term corresponds to the overhead of the overlaps.
This is minimized for $h = \sqrt{n/((k+1)(\ell_0 + 1))}$ and the space becomes 
$O(n)$ as long as 
$k((k+1)(\ell_0 + 1))^{k + 1/2} = O(\sqrt{n})$.

Using this value of $h$, we increase the time complexity from Eq.~\ref{eq:first} to 
\begin{multline}
O(n^{1.5}((k+1)(\ell_0 + 1))^{k-1/2}(\log n - \\ 
\log(\sqrt{n/((k+1)(\ell_0 + 1))}) + k\log \ell_0 + k\log k)(\log k + \log\ell_0)) = \\
O(n^{1.5}((k+1)(\ell_0+1))^{k-1/2}(\log n + k\log \ell_0 + k\log k)(\log k + \log\ell_0)) = \\
O(n^{1.5} (k\ell_0)^{O(k)} \log^2 n).
\end{multline}

This is $O(n^{1.5}\operatorname{polylog}(n))$ for, e.g., 
$k = O(\log\log n)$ and $\ell_0 = \operatorname{polylog}(n)$.
For the case of $\ell_0 = O(k)$, the time complexity remains subquadratic for 
$k = O(\log^{1-\varepsilon} n)$, for any constant $\varepsilon > 0$.

Note that we can reduce this time complexity by using a smaller value of $h$, 
but the space will remain linear only if the correspondingly stricter requirements on 
$k$ and $\ell_0$ are fulfilled.

\subsection{Strided diagonal-wise scan over the matrix $\phi$} 
%%%%%%%%%%%%%%%%%%%%%%%%%%%%%%%%%%%%%%%%%%%%%%%%%%%%

This algorithm is a refinement of the simple technique 
by Flouri et al.~\cite[Sect.~4]{FGKU2014}.

The function $\phi(i, j)$ was defined in Introduction.
Flouri et al. consider a conceptual matrix with the $\phi$ values 
and scan (compute) it diagonal-wise, e.g., after $\phi(3, 1)$ the next 
computed values are: $\phi(4, 2)$, $\phi(5, 3)$, \ldots, $\phi(n, n-2)$.
The desired $\ell_k$ value is the maximum among the computed cells 
and it can be found using constant space (apart from the input sequences themselves), 
while the time complexity is $O(n^2)$.

We reduce the time complexity of the cited technique by factor 
$\ell_k / k$, 
yet the price we pay is $O(n)$ extra space. 
This space is spent for two LCA structures (based on a generalized suffix tree), 
which allow to answer longest common extension (LCE) queries 
in constant time~\cite{BFC2000}.
One of these structures works on the concatenation $S_1\#S_2$, 
where $\#$ is a unique symbol, 
and the other on the same sequence reversed.
Thanks to these structures, we can get in constant time the longest exact match 
starting (resp. ending) at $S_1[i]$ and $S_2[j]$, for any $i$ and $j$. 

Now we can present the algorithm, which is very simple.  
Like Flouri et al., we scan the diagonals, but with two modifications:
$(i)$ we do it a multiple number of times,
$(ii)$ in each pass we visit every $h_i$-th cell, 
where the exact value of $h_i$ for $i$-th pass will be given later.
For each visited cell we compute $k+1$ matching pieces (with single mismatches between) 
going forward along the diagonal, and similarly going backward along it.
From the $2k+1$ possible candidates
we choose the longest match with $k$ mismatches 
involving the currently visited cell. 
Thanks to the LCE-answering data structures we do it in $O(k)$ time.
Note that if $h_i \leq \ell_k$, then visiting the diagonals in strides 
of $h_i$ cells cannot miss a longest common substring with $k$ mismatches.
We start with $h_1 = \min((k+1) \ell_0 + k, n)$ and if we find a common substring 
of length $\geq h_1$ (which may be only equal to $h_1$ in this case), then we stop, 
as we must have found a longest common substring.
If not, we set $h_2 = h_1 / 2$ and scan the matrix again, etc.
The $i$-th iteration is the last one whenever a common substring of length $\geq h_i$ 
is found. 
Note that $h_i > \ell_k / 2$ if $i$-th is the last iteration, 
since otherwise $h_{i-1} \leq \ell_k$ and a longest 
common substring could not have been missed in the previous pass.
In this way, summing a geometric series, 
we immediately obtain the $O(n^2 k / \ell_k)$ time complexity.

Note that the LCE queries are also used by Flouri et al., in another variant 
of their technique, 
but with $O(n + |\mathcal{K}|)$ time (and also $O(n)$ space), 
where $\mathcal{K} = \{(i, j): S_1[i] \neq S_2[j]\}$, which seems to be ``typically'' 
worse.
Yet, the worst case time for both algorithms is quadratic.

\subsection{Faster diagonal processing with tabulation} 
%%%%%%%%%%%%%%%%%%%%%%%%%%%%%%%%%%%%%%%%%%%%%%%%%%%%

Again we work on the technique by Flouri et al.
Recall the assumption that we have a machine word of length $w = \Theta(\log n)$. 
We define an $(f)$-word as a machine word logically divided into 
$\lfloor w/f \rfloor$ fields of $f$ bits. 
Given an $(f)$-word $W$, we denote with $W[i]$ its $i$-th field, 
for $i \in \{1, \ldots, \lfloor w/f \rfloor\}$.
First we pack the sequences $S_1$ and $S_2$, so that each symbol 
is stored in $\lceil \log\sigma \rceil$ bits. 
This step takes linear time and space.
W.l.o.g. assume that $\log\sigma$ is an integer.
Each word of the packed representation will thus 
store $\Theta(\log n / \log\sigma)$ symbols.
%%, where $f = \log\sigma \rceil$.
%% and can be logically divided into fields of $f$ bits.

Additionally we build two lookup tables, $L_1$ and $L_2$. 
The input of $L_1$ is a bit-vector of size $b$ and an integer $0 \leq k' \leq b$, 
and it returns the start and the end position of the largest contiguous area 
of the bit-vector containing at most $k'$ set bits, 
and the number $k'' \leq k'$ of the set bits in the returned area.
$L_2$ works similarly, but it accepts two bit-vectors of size $b$ instead of one, 
$k'$ is upper-bounded by $2b$, 
and the returned area must comprise a (possibly empty) suffix of the first bit-vector 
and a (possibly empty) prefix of the second bit-vector.
More formally, 
$L_1(B[1 \ldots b], k') = (1, b, k'')$ if $popc(B[1 \ldots b]) = k'' < k'$,
and 
$L_1(B[1 \ldots b], k') = (i, j, k')$ if $popc(B[i \ldots j]) = k'$,
and there is no pair of indices $1 \leq i' \leq j' \leq b$ such that 
$j' - i' > j - i$ and $popc(B[i' \ldots j']) = k'$.
Analogously, 
$L_2(B_1[1 \ldots b], B_2[1 \ldots b], k') = (1, 2b, k'')$ 
if $popc(B_1[1 \ldots b] B_2[1 \ldots b]) = k'' < k'$,
and 
$L_2(B_1[1 \ldots b], B_2[1 \ldots b], k') = (i, j, k')$, $i \leq b + 1$, 
$j \geq b$, 
if $popc(B_1[i \ldots b] B_2[1 \ldots j-b]) = k'$, 
and there is no pair of indices $1 \leq i' \leq b+1$, $b \leq j' \leq 2b$, such that 
$j' - i' > j - i$ and $popc(B_1[i' \ldots b] B_2[1 \ldots j-b]) = k'$.

Note that $L_1$ and $L_2$ can be (na{\"i}vely) built in $O(2^{2b} b^3)$ time 
(for all possible inputs, including all possible values of $k'$)
and require $O(2^{2b} b)$ words of space.
We set $b = \log n / 3$ and the LUT construction costs become $o(n)$.

Now we consider all $n$ alignments of sequence $S_2$ against $S_1$, 
that is, for $j$-th alignment, $0 \leq j < n$, 
we look for the longest common substring with $k$ mismatches 
starting at symbols $S_1[i]$ and $S_2[i+j]$, correspondingly, for all valid $i$.
For each alignment, we produce a sequence $W_1, \ldots, W_{n\log\sigma/\log n}$
of $(\log\sigma)$-words 
such that $W_{i'}[j'] = 0$ iff 
$S_1[i'(\log n/\log\sigma)+j'] = S_2[i'(\log n/\log\sigma)+j'+j]$ 
and $2^{\log\sigma-1}$ otherwise.
This can be achieved in $O(n\log\sigma/\log n)$ time 
with the solution from~\cite[Sect.~4]{BreslauerGG2012} 
or a simpler one from~\cite[Sect.~3]{GGF2013} 
(using the primitive $\textsf{fnf}(A)$).

%% Then we work in blocks of $b = \Theta(\log n / \log\sigma)$ symbols.
Each resulting $(\log\sigma)$-word may contain the beginning of an $k$-LCF, 
hence we use $L_1$ for the successive $W_{i'}$ words 
and $L_2$ for substrings starting in $W_{i'}$ and ending in $W_{i'+h}$, 
for all valid $i'$ and $h \geq 1$.
The number of set bits in $W_{i'+1}, \ldots, W_{i'+h-1}$ is obtained 
incrementally, with aid of $L_1$.
In this way we process each alignment in $O(n\log\sigma/\log n)$ time 
and the overall time complexity becomes $O(n^2\log\sigma/\log n)$, 
with linear space.

We can refine the described technique slightly, replacing the $\log\sigma$ 
factor with $\log\min(k+\ell_0, \sigma)$.
To this end, we divide $S_1$ and $S_2$ into substrings of length 
$2((k+1)\ell_0 + k) - 1$, with overlaps of length $(k+1)\ell_0 + k$.
%% \footnote{Naturally, we adapt a well-known trick of working in overlapping 
%% blocks, see, e.g., ...}.
For each such pair of substrings, one from $S_1$ and the other from $S_2$, 
we perform alphabet remapping into the set of symbols occurring in these 
substrings, 
which gives a denser representation of the two sequences 
if $\log k + \log\ell_0 = \Theta(\log(k + \ell_0)) = o(\log\sigma)$.
This preprocessing takes $O(n^2\log(k\ell_0)/(k\ell_0)^2)$
%%(\ell_0+1)^2 k^2 
%% \log((\ell_0+1)k))$ 
overall time, 
where the logarithmic factor comes from the BST operations 
in alphabet remapping.
%% Note that $(k+1)\ell_0 + k \geq \ell_k$, which means 
%% that 
As $(k+1)\ell_0 + k \geq \ell_k$,
an $k$-LCF will be found in at least one pair of our overlapping substrings, 
in $O(n^2\log\min(k+\ell_0, \sigma)/\log n)$ time.
Note however that the preprocessing time
%% due to alphabet mapping, 
%% i.e., the additive term $O(n^2\log(k\ell_0)/(k\ell_0)^2)$, 
may be dominating for small $k$ and $\ell_0$.
On the other hand, it is easy to check that for (e.g.)
%% $k\ell_0 = O(\log n)$ 
$k\ell_0 = O(\log^{1-\varepsilon} n)$, for any constant $\varepsilon > 0$,
%% one of the two previously presented algorithms 
the algorithm from Subsect.~\ref{sec:ng}
works in $o(n^2/\log n)$ time, unreachable for the presented tabulation technique.
One can also notice that the refinement does not help 
when $\log\sigma = O(\log\log n)$, or, in other words, 
that its total time complexity is 
%% $\Omega(n^2\log\log n/\log n)$ 
%% for $\log\sigma = \Omega(\log\log n)$.
$\Omega(n^2\log\min(\sigma, \log n)/\log n)$.

%% file: concs.tex
\section{Conclusions}

We presented three algorithms for the recently introduced problem of 
finding the longest common substring of two strings with $k$ mismatches.
The first algorithm obtains $O(n^{1.5}\operatorname{polylog}(n))$ time 
for a small $k$ (namely, $k = O(\log^{1-\varepsilon} n)$, for any constant 
$\varepsilon > 0$) 
when $\ell_0 = O(k)$.
The second algorithm obtains $o(n^2)$ time 
when $\ell_k = \omega(k)$; if $\ell_0 = \omega(k)$ then obviously 
$\ell_k = \omega(k)$ is satisfied as well.
The conclusion is thus: 
if $k = O(\log^{1-\varepsilon} n)$, for any constant $\varepsilon > 0$, 
then after a linear time preprocessing in which we find the value of $\ell_0$, 
we can choose one from the two presented algorithms to obtain subquadratic 
overall time complexity, using linear space.
So far, subquadratic time for this problem (namely, $O(n\log n)$) 
was known only for the case of $k = 1$~\cite{FGKU2014}.
The last algorithm, based on tabulation, 
gives another niche for (slightly) subquadratic 
behavior: either if $\log\sigma = o(\log n)$, 
or both $k\ell_0 = \omega(1)$ and $\log(k + \ell_0) = o(\log n)$ hold.

A few questions may be now posed. 
Can the first algorithm be practical for real data and small $k$, 
possibly in a variant without a worst-case guarantee, e.g., employing a Bloom filter?
Can sophisticated solutions for dictionary matching with errors, e.g.~\cite{CGL2004}, 
broaden its range of applicability (in theory or in practice)?
Are bit-parallel techniques promising for the $k$-LCF problem?
Related to the last question, 
it might be possible (although is not obvious how) to apply the techniques 
from~\cite{GGF2013}.